\documentclass[11pt,superscriptaddress,aps,prd,preprint]{revtex4-1}
\usepackage{amsmath}
\usepackage{amssymb}
\usepackage{graphicx}
\usepackage{slashed}

\newcommand{\bea}{\begin{eqnarray}}
\newcommand{\eea}{\end{eqnarray}}

\newcommand{\e}{\epsilon}

\begin{document}

\title{On Fermion-Graviton Scattering at Finite Temperature}

\author{A. F. Santos}
\email{alesandroferreira@fisica.ufmt.br}
\affiliation{Instituto de F\'{\i}sica, Universidade Federal de Mato Grosso,\\
78060-900, Cuiab\'{a}, Mato Grosso, Brazil.}
\affiliation{Department of Physics and Astronomy, University of Victoria,
3800 Finnerty Road Victoria, BC, Canada.}

\author{S. C. Ulhoa }
\email{sc.ulhoa@gmail.com} \affiliation{Instituto de F\'{i}sica,
Universidade de Bras\'{i}lia, 70910-900, Bras\'{i}lia, DF,
Brazil.}

\author{Faqir C. Khanna\footnote{Professor Emeritus - Physics Department, Theoretical Physics Institute, University of Alberta - Canada}}
\email{khannaf@uvic.ca}
\affiliation{Department of Physics and Astronomy, University of Victoria,
3800 Finnerty Road Victoria, BC, Canada.}


\begin{abstract}

The early universe is dominated by phenomena at high temperatures. The thermal effects decrease during evolution of the universe. However there are some phenomena, such as processes inside stars and black holes, where the role of the temperature is important. Study of processes involving gravitons at finite temperature is a viable means to understand gravitational processes in the universe. Using the Thermo Field Dynamics formalism transition amplitudes involving gravitons and fermions at finite temperature are calculated. In this approach our results consist of two parts: one corresponding to the zero temperature part and the other corresponding to a temperature dependent part.

\end{abstract}


\maketitle
\section{Introduction}

In the beginning of the Universe gravity exhibits quantum features at extremely high temperatures. For instance during the inflation epoch it is possible that time and temperature worked as entangled quantities~\cite{infla}. The quantum effects at high temperature have been analyzed in Kaluza-Klein cosmology \cite{htkk}. The classical theories of gravity, Einstein (general) theory and teleparallel gravity, have provided in depth understanding of the universe at zero temperature. The teleparallel gravity, formulated in terms of torsion, is an alternative theory of gravitation and dynamically equivalent to Einstein relativity, in terms of the curvature of space \cite{Einst, Maluf}. The difference between them resides in the definition of gravitational energy and angular momentum. It is interesting to note that the graviton propagator obtained from the teleparallel Lagrangian density is different from the one generated in general relativity. Recently the scattering of gravitons and fermions in the framework of teleparallel gravity~\cite{sfg} was considered. However still finite temperature needs to be incorporated in these scattering processes.

In order to treat the dynamics of systems properly temperature has to be incorporated. Temperature effects may be introduced by two equivalent methods: (i) the imaginary time formalism or the Matsubara formalism \cite{Matsubara} and (ii) the real time formalisms. There are two distinct, but equivalent methods: (a) the closed time path formalism \cite{Schwinger} and (b) the Thermo Field Dynamics (TFD) method \cite{Umezawa1, Umezawa2, Umezawa22, Khanna1, Khanna2}. Here the TFD method is chosen. In this formalism the expectation value of a physical observable $A$ in a thermal vacuum, $|0(\beta) \rangle$, is interpreted as the statistical average, i.e., $\langle 0(\beta)| A|0(\beta) \rangle=\langle A \rangle$, where $\beta=\frac{1}{k_BT}$ with $T$ being the temperature and $k_B$ the Boltzmann constant. In order to satisfy this relation between expectation value of $A$ in a thermal vacuum and statistical average, $\langle A \rangle$, is necessary to extend the usual Hilbert space $S$ to $S_T=S\otimes\tilde{S}$, where $\tilde{S}$ is a fictitious Hilbert space. These two spaces are connected by Bogoliubov transformations, which introduce temperature effects. The Bogoliubov transformation introduces a rotation in the tilde and nontilde operators. In addition, the propagators in TFD are written as the sum of two parts:  one part of the propagator is zero temperature while the other corresponds to the finite temperature.

The paper is structured in the following form. In section II, the Lagrangian density describes a free fermion and graviton field and the interaction between them. In section III, the Thermo Field Dynamics is introduced. In sections IV and V, the graviton and fermion propagators at finite temperature are presented. In section VI, the interaction between gravitons and fermions at finite temperature is analyzed. Finally in the last section some concluding remarks are presented.

\section{Lagrangian Density}\label{ld}

In this section the lagrangian density of the interacting teleparallel gravity and fermionic field is presented. Here gravitation is described by teleparallelism formalism. Such a theory is formulated in terms of the tetrad field, which is related to the metric, $g_{\mu\nu}=e^a\,_\mu e_{a\nu}$, in a space-time with torsion. This space-time is called Weitzenb\"ock space which is endowed with Cartan connection, $\Gamma_{\mu \lambda\nu}=e_{a\mu}\partial_{\lambda} e^{a}\,_{\nu}$. The skew symmetric part of the torsion is

\begin{equation}
T^{a}\,_{\lambda\nu}=\partial_{\lambda} e^{a}\,_{\nu}-\partial_{\nu}
e^{a}\,_{\lambda}\,. \label{3}
\end{equation}
In the tetrad field latin indices, $a$, represent the Lorentz group and greek indices, $\mu$, indicate the transformation of coordinates. The Cartan connection satisfies the identity
\begin{equation}
\Gamma_{\mu \lambda\nu}= {}^0\Gamma_{\mu \lambda\nu}+ K_{\mu
\lambda\nu}\,, \label{2}
\end{equation}
where

\begin{eqnarray}
K_{\mu\lambda\nu}&=&\frac{1}{2}(T_{\lambda\mu\nu}+T_{\nu\lambda\mu}+T_{\mu\lambda\nu})\label{3.5}
\end{eqnarray}
is the contortion tensor. Now it is possible to relate the torsion tensor to the scalar curvature defined in a Riemann space. As a consequence the Hilbert-Einstein Lagrangian density assumes the following form in Weitzenb\"ock space

\bea
eR({}^0\Gamma)\equiv -e(\frac{1}{4}T^{abc}T_{abc}+\frac{1}{2}T^{abc}T_{bac}-T^aT_a)+2\partial_\mu(eT^\mu)\,,\label{5}
\eea
where $e$ is the determinant of the tetrad field, $T_a=T^b\,_{ba}$ and $T_{abc}=e_b\,^\mu e_c\,^\nu T_{a\mu\nu}$. If the last term (total divergence) is dropped, since it does not affect the field equation, then the gravitational Lagrangian density is
\bea
{\cal L}_G= -k e(\frac{1}{4}T^{abc}T_{abc}+\frac{1}{2}T^{abc}T_{bac}- T^aT_a) \label{lag}\,,
\eea
where $k=1/16\pi$. It can be rewritten as
\begin{equation}
{\cal L}_G\equiv -ke\Sigma^{abc}T_{abc} \,,
\label{5.1}
\end{equation}
where
\bea
\Sigma^{abc}=\frac{1}{4} (T^{abc}+T^{bac}-T^{cab}) +\frac{1}{2}(
\eta^{ac}T^b-\eta^{ab}T^c)\,. \label{6}
\eea
Since this Lagrangian density is equivalent to that of Einstein relativity: then both theories are dynamically equivalent. Hence this Lagrangian density $({\cal L}_G)$ is used to calculate all processes related to gravitons.

The complete Lagrangian density that describes the interaction between gravitons and fermions is given by
\bea
{\cal L}={\cal L}_G+{\cal L}_M,
\eea
where ${\cal L}_M$ stands for the fermion field coupled to gravity. It is given by
\bea
{\cal L}_M=e\left[\frac{i}{2}\left(\bar{\psi}\gamma^\mu D_\mu\psi-D_\mu\bar{\psi}\gamma^\mu\psi\right)-m\bar{\psi}\psi\right]\,,
\eea
with $D_\mu=\partial_\mu+\frac{i}{4}\,K_{\mu ab}\Sigma^{ab}$ and  $\Sigma^{ab}=\frac{i}{2}[\gamma^a,\gamma^b]$, $\psi$ is the Fermi field with mass $m$, $\bar{\psi}=\psi^\dagger\gamma_0$ and $\gamma^\mu$ are Dirac matrices. Thus the fermionic Lagrangian density is separated into two parts as ${\cal L}_M={\cal L}_F+{\cal L}_{GF}$, where
\bea
{\cal L}_F=e\left[\frac{i}{2}\left(\bar{\psi}\gamma^\mu \partial_\mu\psi-\partial_\mu\bar{\psi}\gamma^\mu\psi\right)-m\bar{\psi}\psi\right]\,,
\eea
stands for the free fermion field and
\bea
{\cal L}_{GF}=-\frac{e}{8}\left(\bar{\psi}\gamma^\mu K_{\mu ab}\Sigma^{ab}\psi+\bar{\psi}K_{\mu ab}\Sigma^{ab}\gamma^\mu\psi\right)\,,
\eea
describes the interaction between fermions and gravitons.

\section{Thermo Field Dynamics - TFD}\label{tfd}

Here a brief introduction to TFD formalism is presented. This formalism is constructed with two ingredients: (i) a doubling of the Fock space, $S$, giving rise to $ S_T=S\otimes \tilde S$, applicable to equilibrium systems. The physical quantities are described by the nontilde operators. (ii) Bogoliubov transformation introduces thermal effects through a rotation in the tilde and nontilde operators. The Bogoliubov transformations for bosons and fermions are distinct.

\subsection{For bosons}

Bogoliubov transformations for bosons are
\bea
A_{a\mu}(k)&=&c_BA_{a\mu}(k,\beta)+d_B\tilde A_{a\mu}^\dagger(k,\beta),\nonumber\\
A_{a\mu}^\dagger(k)&=&c_BA_{a\mu}^\dagger(k,\beta)+d_B\tilde A_{a\mu}(k,\beta),\nonumber\\
\tilde A_{a\mu}(k)&=&c_B\tilde A_{a\mu}(k,\beta)+d_B A_{a\mu}^\dagger(k,\beta),\nonumber\\
\tilde A_{a\mu}^\dagger(k)&=&c_B\tilde A_{a\mu}^\dagger(k,\beta)+d_B A_{a\mu}(k,\beta),\label{BTp}
\eea
where $(A_{a\mu}^\dagger, \tilde A_{a\mu}^\dagger)$ are creation operators, $(A_{a\mu}, \tilde A_{a\mu})$ are destruction operators and
\bea
c_B^2=1+f_B(\omega),\quad\quad d_B^2=f_B(\omega), \quad\quad f_B(\omega)=\frac{1}{e^{\beta\omega}-1}.\label{phdef}
\eea
Here $\omega=\omega(k)$. These thermal operators satisfy the algebraic relations
\bea
\left[A_{a\mu}(k, \beta), A_{b\nu}^\dagger(p, \beta)\right]&=&\delta^3(k-p)\eta_{ab}\,g_{\mu\nu},\nonumber\\
\left[\tilde A_{a\mu}(k, \beta), \tilde A_{b\nu}^\dagger(p, \beta)\right]&=&\delta^3(k-p)\eta_{ab}\,g_{\mu\nu},\label{ComB}
\eea
and other commutation relations are null.

\subsection{For fermions}

Bogoliubov transformations for fermions are
\bea
A_{a\mu}(k)&=&c_FA_{a\mu}(k,\beta)+d_F\tilde A_{a\mu}^\dagger(k,\beta),\nonumber\\
A_{a\mu}^\dagger(k)&=&c_F A_{a\mu}^\dagger(k,\beta)+d_F \tilde A_{a\mu}(k,\beta),\nonumber\\
\tilde A_{a\mu}(k)&=&c_F\tilde A_{a\mu}(k,\beta)-d_F A_{a\mu}^\dagger(k,\beta),\nonumber\\
\tilde A_{a\mu}^\dagger(k)&=&c_F\tilde A_{a\mu}^\dagger(k,\beta)-d_F A_{a\mu}(k,\beta),\label{BTf}
\eea
with
\bea
c_F^2=1-f_F(\omega),\quad\quad d_F^2=f_F(\omega), \quad\quad f_F(\omega)=\frac{1}{e^{\beta\omega}+1}.\label{ferdef}
\eea
In this case the algebraic relations are
\bea
\left\{A_{a\mu}(k, \beta), A_{b\nu}^\dagger(p, \beta)\right\}&=&\delta^3(k-p)\eta_{ab}\,g_{\mu\nu},\nonumber\\
\left\{\tilde A_{a\mu}(k, \beta), \tilde A_{b\nu}^\dagger(p, \beta)\right\}&=&\delta^3(k-p)\eta_{ab}\,g_{\mu\nu}.\label{ComF}
\eea
Other commutation relations are null.

\section{The Graviton Propagator at finite temperature}\label{gp}

The graviton propagator in weak field approximation is defined as
\bea
iD^{kl}_{ab\mu\nu}(x-y)=\left\langle 0(\beta)\left|\tau\left[e^k_{a\mu}(x)e^l_{b\nu}(y)\right]\right|0(\beta)\right\rangle,
\eea
where $\tau$ is the time ordering operator, $k,l=1,2$ and $a,b,\mu,\nu$ are tensor indices. In this approximation $g_{\mu\nu}=\eta_{\mu\nu}+h_{\mu\nu}$. In teleparallel gravity the fundamental tetrad field is $e_{a\mu}(x)$. A Fourier expansion of this field is written as
\bea
e_{a\mu}(x)=\int\frac{d^3k}{\sqrt{2\omega_k(2\pi)^3}}\left(A_{a\mu}(k)e^{-ik_\sigma x^\sigma}+A_{a\mu}^\dagger(k)e^{ik_\sigma x^\sigma}\right),\label{Fund}
\eea
where $A_{a\mu}(k)$ and $A_{a\mu}^\dagger(k)$ are the annihilation and creation operators, respectively.

The propagator in TFD is a thermal doublet and has 2$\times$2 matrix structure. Using the thermal doublet notation,
\bea
\left( \begin{array}{cc} e_{a\mu}^1 \\
e_{a\mu}^2 \end{array} \right) = \left( \begin{array}{cc} e_{a\mu} \\
\tilde{e}_{a\mu}^\dagger \end{array} \right),
\eea
the component $k=l=1$ is written explicitly as
\bea
iD^{11}_{ab\mu\nu}(x-y)&=&\theta(t_x-t_y)\left\langle 0(\beta)\left|\left[e_{a\mu}(x)e_{b\nu}(y)\right]\right|0(\beta)\right\rangle\nonumber\\
&+&\theta(t_y-t_x)\left\langle 0(\beta)\left|\left[e_{b\nu}(y)e_{a\mu}(x)\right]\right|0(\beta)\right\rangle,\label{Prop1}
\eea
where $\theta(t_x-t_y)$ is the step function. For simplicity, the components are calculated separately. Substituting the fundamental field, eq. (\ref{Fund}), the 11-component of the propagator becomes
\bea
iD^{11}_{ab\mu\nu}(x-y)&=&\theta(t_x-t_y)\int\frac{d^3k}{\sqrt{2\omega_k(2\pi)^3}}\int\frac{d^3p}{\sqrt{2\omega_p(2\pi)^3}}\times\\
&\times &\langle 0(\beta)|(A_{a\mu}(k)e^{-ik_\sigma x^\sigma}+A_{a\mu}^\dagger(k)e^{ik_\sigma x^\sigma})(A_{b\nu}(p)e^{-ip_\sigma y^\sigma}+A_{b\nu}^\dagger(p)e^{ip_\sigma y^\sigma})|0(\beta)\rangle\nonumber\\
&+&\theta(t_y-t_x)\int\frac{d^3p}{\sqrt{2\omega_p(2\pi)^3}}\int\frac{d^3k}{\sqrt{2\omega_k(2\pi)^3}}\times\nonumber\\
&\times &\langle 0(\beta)|(A_{b\nu}(p)e^{-ip_\sigma y^\sigma}+A_{b\nu}^\dagger(p)e^{ip_\sigma y^\sigma})(A_{a\mu}(k)e^{-ik_\sigma x^\sigma}+A_{a\mu}^\dagger(k)e^{ik_\sigma x^\sigma})|0(\beta)\rangle.\nonumber
\eea
Using the Bogoliubov transformations, eqs. (\ref{BTp}), we get
\bea
iD^{11}_{ab\mu\nu}(x-y)&=&\theta(t_x-t_y)\int\frac{d^3k}{\sqrt{2\omega_k(2\pi)^3}}\int\frac{d^3p}{\sqrt{2\omega_p(2\pi)^3}}\times\\
&\times &\langle 0(\beta)|[c_B^2A_{a\mu}(k)A_{b\nu}^\dagger(p)e^{-ik_\sigma x^\sigma+ip_\sigma y^\sigma}+d_B^2\tilde{A}_{a\mu}(k)\tilde{A}_{b\nu}^\dagger(p)e^{ik_\sigma x^\sigma-ip_\sigma y^\sigma}]|0(\beta)\rangle\nonumber\\
&+&\theta(t_y-t_x)\int\frac{d^3p}{\sqrt{2\omega_p(2\pi)^3}}\int\frac{d^3k}{\sqrt{2\omega_k(2\pi)^3}}\times\nonumber\\
&\times &\langle 0(\beta)|[c_B^2A_{b\nu}(p)A_{a\mu}^\dagger(k)e^{-ip_\sigma y^\sigma+ik_\sigma x^\sigma}+d_B^2\tilde{A}_{b\nu}(p)\tilde{A}_{a\mu}^\dagger(k)e^{ip_\sigma y^\sigma-ik_\sigma x^\sigma}]|0(\beta)\rangle\nonumber.
\eea
Assuming that commutation relations, eqs. (\ref{ComB}), are satisfied, this component of the graviton propagator becomes
\bea
iD^{11}_{ab\mu\nu}(x-y)&=&-\int\frac{d^3k}{(2\pi)^3}\frac{1}{2\omega_k}\eta_{ab}\,\eta_{\mu\nu}\Biggl\{ c_B^2\Bigl[\theta(t_x-t_y)e^{-ik_\sigma(x^\sigma-y^\sigma)}+\theta(t_y-t_x)e^{ik_\sigma(x^\sigma-y^\sigma)}\Bigl]\nonumber\\
&+&d_B^2\Bigl[\theta(t_x-t_y)e^{ik_\sigma(x^\sigma-y^\sigma)}+\theta(t_y-t_x)e^{-ik_\sigma(x^\sigma-y^\sigma)}\Bigl]\Biggl\},
\eea
where the integral over $d^3p$ has been calculated. Using the Cauchy theorem
\bea
\int dk^0\frac{e^{-ik_0(x_0-y_0)}}{k_0-(\omega_k-i\xi)}&=&-2\pi ie^{-i\omega_k(x_0-y_0)}\theta(x_0-y_0),\nonumber\\
\int dk^0\frac{e^{-ik_0(x_0-y_0)}}{k_0-(-\omega_k+i\xi)}&=&2\pi ie^{i\omega_k(x_0-y_0)}\theta(y_0-x_0),\label{Cauchy}
\eea
the 11-component is given as
\bea
D^{11}_{ab\mu\nu}(x-y)&=&-\int\frac{d^4k}{(2\pi)^4}e^{-ik_\sigma(x^\sigma-y^\sigma)}\eta_{ab}\,\eta_{\mu\nu}\left[\frac{c_B^2}{k_0^2-(\omega_k-i\epsilon)^2}-\frac{d_B^2}{k_0^2-(\omega_k+i\epsilon)^2}\right].\label{Comp11}
\eea

Similarly the component $k=1$ and $l=2$ is written as
\bea
iD^{12}_{ab\mu\nu}(x-y)&=&\theta(t_x-t_y)\langle 0(\beta)|[e_{a\mu}(x)\tilde{e}_{b\nu}^\dagger(y)]|0(\beta)\rangle\nonumber\\
&+&\theta(t_y-t_x)\langle 0(\beta)|[\tilde{e}_{b\nu}^\dagger(y)e_{a\mu}(x)]|0(\beta)\rangle.\label{Prop2}
\eea
Following similar steps used for the 11-component, we get
\bea
D^{12}_{ab\mu\nu}(x-y)&=&-\int\frac{d^4k}{(2\pi)^4}e^{-ik_\sigma(x^\sigma-y^\sigma)}\eta_{ab}\,\eta_{\mu\nu}\left[\frac{c_Bd_B}{k_0^2-(\omega_k-i\epsilon)^2}-\frac{c_Bd_B}{k_0^2-(\omega_k+i\epsilon)^2}\right].\label{Comp12}
\eea
Then for the component $k=2$ and $l=1$ we have
\bea
D^{21}_{ab\mu\nu}(x-y)&=&D^{12}_{ab\mu\nu}(x-y).
\eea

Finally the 22-component is
\bea
D^{22}_{ab\mu\nu}(x-y)&=&-\int\frac{d^4k}{(2\pi)^4}e^{-ik_\sigma(x^\sigma-y^\sigma)}\eta_{ab}\,\eta_{\mu\nu}\left[\frac{d_B^2}{k_0^2-(\omega_k-i\epsilon)^2}-\frac{c_B^2}{k_0^2-(\omega_k+i\epsilon)^2}\right].\label{Comp22}
\eea

The graviton propagator at finite temperature is
\bea
D^{kl}_{ab\mu\nu}(x-y)=i\int\frac{d^4k}{(2\pi)^4}e^{-ik_\sigma(x^\sigma -y^\sigma)} \,D^{kl}_{ab\mu\nu}(k),
\eea
where
\bea
-iD^{kl}_{ab\mu\nu}(k)=\left( \begin{array}{cc} \frac{c_B^2}{k_0^2-(\omega_k-i\epsilon)^2}-\frac{d_B^2}{k_0^2-(\omega_k+i\epsilon)^2} \hspace{0,2cm} & \hspace{0,2cm} \frac{c_Bd_B}{k_0^2-(\omega_k-i\epsilon)^2}-\frac{c_Bd_B}{k_0^2-(\omega_k+i\epsilon)^2} \\
\frac{c_Bd_B}{k_0^2-(\omega_k-i\epsilon)^2}-\frac{c_Bd_B}{k_0^2-(\omega_k+i\epsilon)^2} \hspace{0,2cm} & \hspace{0,2cm} \frac{d_B^2}{k_0^2-(\omega_k-i\epsilon)^2}-\frac{c_B^2}{k_0^2-(\omega_k+i\epsilon)^2} \end{array} \right)\,\eta_{ab}\,\eta_{\mu\nu}.\label{4}
\eea

Using relations given in eq. (\ref{phdef}), the propagator is separated as
\bea
-iD^{kl}_{ab\mu\nu}(k)=-iD^{kl}_{(0)ab\mu\nu}(k)-iD^{kl}_{(\beta)ab\mu\nu}(k),
\eea
where $D^{kl}_{(0)ab\mu\nu}(k)$ and $D^{kl}_{(\beta)ab\mu\nu}(k)$ are zero and finite temperature components respectively. Explicitly
\bea
-iD^{kl}_{(0)ab\mu\nu}(k)=\eta_{ab}\,\eta_{\mu\nu}\left[k_0^2-(\omega_k-i\epsilon\xi)^2\right]^{-1}\xi\label{Gzero}
\eea
and
\bea
-iD^{kl}_{(\beta)ab\mu\nu}(k)=-\eta_{ab}\,\eta_{\mu\nu}\frac{2\pi i\delta(k_0^2-\omega_k^2)}{e^{\beta \omega_k}-1}\left( \begin{array}{cc}1&e^{\beta \omega_k/2}\\e^{\beta \omega_k/2}&1\end{array} \right),\label{Gtemp}
\eea
where
\bea
\xi=\left( \begin{array}{cc}1&0\\0&-1\end{array} \right).
\eea
The 11-component yields the physical graviton propagator at zero temperature as
\bea
D_{ab\mu\nu}(k)=\frac{i\eta_{ab}}{k^\mu k^\nu}.
\eea

\section{The Electron Propagator at Finite temperature}\label{ep}

The electron propagator is defined as
\bea
S^{kl}_{\mu\nu}(x-y)&=&\theta(t_x-t_y)\langle 0(\beta)|\psi^k_{\mu}(x)\bar{\psi}^l_{\nu}(y)|0(\beta)\rangle\nonumber\\
& -& \theta(t_y-t_x)\langle 0(\beta)|\bar{\psi}^l_{\nu}(y)\psi^k_{\mu}(x)|0(\beta)\rangle,\label{electron2}
\eea
where $\bar{\psi}(y)=\psi^\dagger(y)\gamma^0$ and $\mu, \nu$ are the spinor indices. Using
\bea
\psi(x)=\sum_r\int d^3k\, \left[u^r(k)a^r(k)e^{i(k\cdot x-\xi_k t)}+v^r(k)b^{r\dagger}(k)e^{-i(k\cdot x-\xi_k t)}\right],
\eea
where $a^r(k)$ and $b^{r}(k)$ are annihilation operators for electrons and positrons, respectively, and Dirac spinors are $u^r(k)$ and $v^r(k)$. Then the electron propagator at finite temperature becomes \cite{AF}
\bea
S^{ab}_{\mu\nu}(k)&=&\left(\frac{\gamma^0\xi-\vec{\gamma}\cdot \vec{k}+m}{2\xi}\right)_{\mu\nu}\left[U_F(\xi)(k_0-\xi +i\delta\tau)^{-1}U_F^\dagger(\xi)\right]^{ab}\nonumber\\&+&\left(\frac{\gamma^0\xi+\vec{\gamma}\cdot \vec{k}-m}{2\xi}\right)_{\mu\nu}\left[U_F(-\xi)(k_0+\xi +i\delta\tau)^{-1}U_F^\dagger(-\xi)\right]^{ab},
\eea
where Bogoliubov transformations, eqs. (\ref{BTf}), are used. Here $\xi\equiv\xi(k)$ and
\bea
U_F(\xi)=\left( \begin{array}{cc}c_F(\xi) & d_F(\xi) \\
-d_F(\xi) & c_F(\xi)\end{array} \right),
\eea
with $c_F(\xi)$ and $d_F(\xi)$ are given by eq. (\ref{ferdef}).

The electron propagator is separated into two parts as
\bea
S_{\mu\nu}(k)=S_{\mu\nu}^{(0)}(k)+S_{\mu\nu}^{(\beta)}(k),\label{fermion}
\eea
where $S_{\mu\nu}^{(0)}(k)$ and $S_{\mu\nu}^{(\beta)}(k)$ are zero and finite temperature parts respectively. Explicitly
\bea
-iS_{\mu\nu}^{(0)}(k)&=&\frac{\slashed k +m}{k^2-m^2},\\
-iS_{\mu\nu}^{(\beta)}(k)&=&\frac{2\pi i}{e^{\beta k_0}+1}\Biggl[\frac{\gamma^0\epsilon-\vec{\gamma}\cdot\vec{k}+m}{2\epsilon}\left( \begin{array}{cc}1&e^{\beta k_0/2}\\e^{\beta k_0/2}&-1\end{array} \right)\delta(k_0-\epsilon)\nonumber\\
&+&\frac{\gamma^0\epsilon+\vec{\gamma}\cdot\vec{k}-m}{2\epsilon}\left( \begin{array}{cc}-1&e^{\beta k_0/2}\\e^{\beta k_0/2}&1\end{array} \right)\delta(k_0+\epsilon)\Biggl].
\eea

\section{Gravitons-Fermions interactions}\label{interaction}

In this section transition amplitudes, ${\cal M}$, for the M$\varnothing$ller scattering, the Compton scattering and a new gravitational scattering process are calculated.

\subsection{Propagators}

The graviton and fermion propagators are represented in FIG.1 and FIG.2, respectively.

\begin{figure}[h]
\includegraphics[scale=0.4]{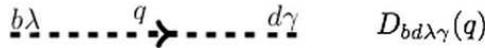}
\caption{ Graviton propagator}
\end{figure}

\begin{figure}[h]
\includegraphics[scale=0.4]{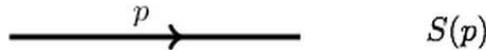}
\caption{ Electron propagator}
\end{figure}

\subsection{Vertex factors}

The vertex of the interaction between two gravitons and two fermions is
\bea
V^{ab\sigma\nu}=\frac{ie}{8}\Sigma^{ab}\eta^{\sigma\nu}\left(\gamma^\mu q_{2\mu}+q_{2\mu}\gamma^\mu\right),
\eea
and is represented in FIG.3.
\begin{figure}[h]
\includegraphics[scale=0.5]{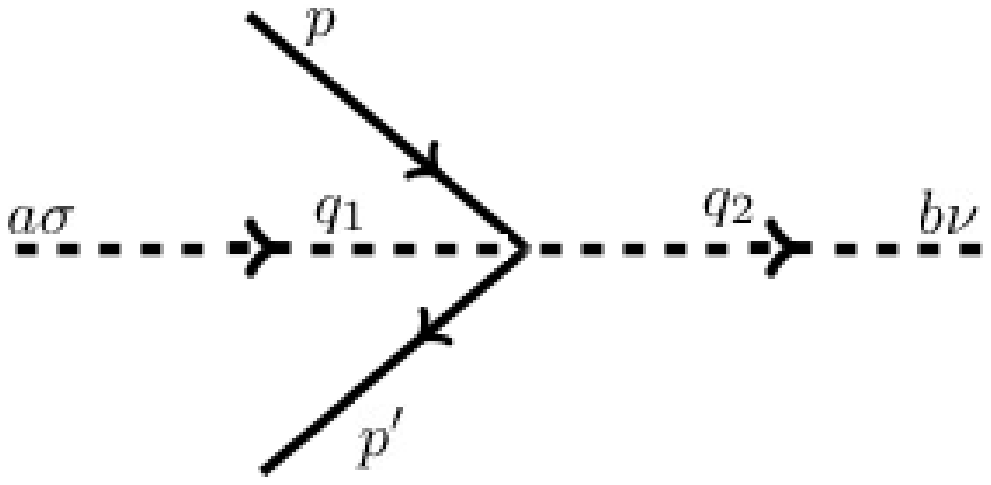}
\caption{Graviton-Fermion Vertex}
\end{figure}

The fermion-graviton vertex is
\bea
{\cal V}^{b\sigma}=\frac{ie}{4}\Sigma^{b\sigma}\left(\gamma^\mu q_\mu+q_\mu\gamma^\mu\right),
\eea
where the approximation $e_{a\nu}=\delta_{a\nu}+\phi_{a\nu}$, with $\delta_{a\nu}$ being the Kronecker delta and a weak field approximation for $\phi_{a\nu}$ is used \cite{sfg}. This vertex is represented in FIG.4.
\begin{figure}[h]
\includegraphics[scale=0.5]{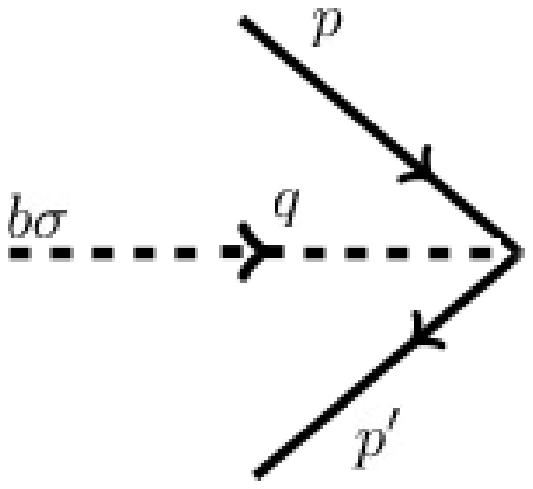}
\caption{Graviton-Fermion Vertex}
\end{figure}

\subsection{Transition amplitudes at finite temperature}

\subsubsection{M$\varnothing$ller scattering}

The M$\varnothing$ller scattering process is described in FIG. 5.
\begin{figure}[h]
\includegraphics[scale=0.5]{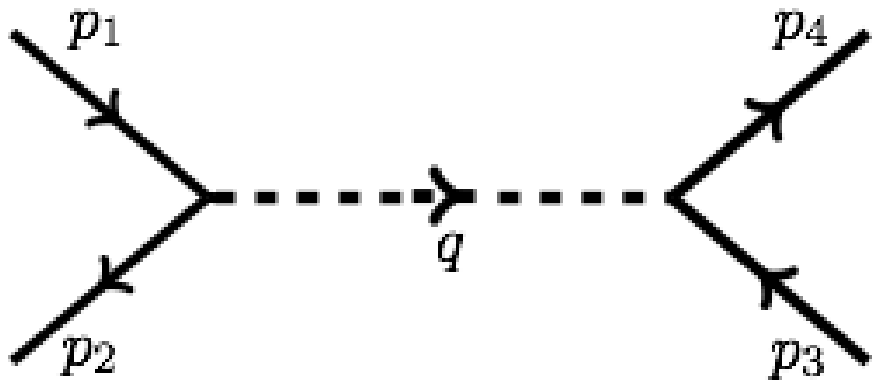}
\caption{ M$\varnothing$ller scattering}
\end{figure}

The transition amplitude is
\begin{equation}
-i{\cal M}=\bar{u}(p_2){\cal V}^{b\sigma}u(p_1)D_{bd\sigma\gamma}(q)\bar{u}(p_4){\cal V}^{d\gamma}u(p_3),
\end{equation}
where ${\cal V}^{b\sigma}$ is the vertex and $D_{bd\sigma\gamma}(q)$ is the graviton propagator at finite temperature. The transition amplitude for this scattering consist of two parts
\bea
{\cal M}={\cal M}_{(0)}+{\cal M}_{(\beta)},
\eea
where ${\cal M}_{(0)}$ and ${\cal M}_{(\beta)}$ are zero and finite temperature components respectively. Zero temperature transition amplitude is
\bea
{\cal M}_{(0)}&=&\frac{e^2}{16}\bar{u}(p_2)u(p_1)\left(\gamma^\mu q_{\mu}+q_{\mu}\gamma^\mu\right)\Sigma_d\,^\sigma\Sigma^d\,_\sigma\times\nonumber\\
&\times &\left[\frac{\xi}{q_0^2-(\omega_q-i\epsilon\xi)^2}\right]\left(\gamma^\nu q_{\nu}+q_{\nu}\gamma^\nu\right)\bar{u}(p_4)u(p_3),
\eea
and the finite temperature part is
\bea
{\cal M}_{(\beta)}&=&\frac{e^2}{16}\bar{u}(p_2)u(p_1)\left(\gamma^\mu q_{\mu}+q_{\mu}\gamma^\mu\right)\Sigma_d\,^\sigma\Sigma^d\,_\sigma\times\nonumber\\
&\times &\left[\frac{2\pi i\delta(q_0^2-\omega_q^2)}{e^{\beta \omega_q}-1}\left( \begin{array}{cc}1&e^{\beta \omega_q/2}\\e^{\beta \omega_q/2}&1\end{array} \right)\right]\left(\gamma^\nu q_{\nu}+q_{\nu}\gamma^\nu\right)\bar{u}(p_4)u(p_3),
\eea
where $\Sigma_d\,^\sigma\Sigma^d\,_\sigma=\eta_{bd}\,\eta_{\sigma\gamma}\Sigma^{d\sigma}\Sigma^{d\gamma}$.

\subsubsection{Compton scattering}

The Compton scattering process is given in FIG. 6.
\begin{figure}[h]
\includegraphics[scale=0.5]{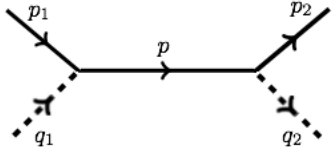}
\caption{Compton scattering}
\end{figure}

The transition amplitude is
\begin{equation}
-i{\cal M}=\bar{u}(p_2){\cal V}^{b\sigma}S(p)\eta_{bc}\,\eta_{\sigma\nu}{\cal V}^{c\mu}u(p_1),
\end{equation}
then
\bea
{\cal M}&=&\frac{e^2}{16}\bar{u}(p_2)\Sigma^{b\sigma}\left(\gamma^\mu q_{1\mu}+q_{1\mu}\gamma^\mu\right)\Sigma_{b\sigma}\left(\gamma^\alpha q_{2\alpha}+q_{2\alpha}\gamma^\alpha\right)u(p_1)\times\nonumber\\
&\times &\Biggl\{\left(\frac{\slashed p+m}{p^2-m^2}\right)+\frac{2\pi i}{e^{\beta p_0}+1}\Biggl[\frac{\gamma^0\epsilon-\vec{\gamma}\cdot\vec{p}+m}{2\epsilon}\left( \begin{array}{cc}1&e^{\beta p_0/2}\\e^{\beta p_0/2}&-1\end{array} \right)\delta(p_0-\epsilon)\nonumber\\
&+&\frac{\gamma^0\epsilon+\vec{\gamma}\cdot\vec{p}-m}{2\epsilon}\left( \begin{array}{cc}-1&e^{\beta p_0/2}\\e^{\beta p_0/2}&1\end{array} \right)\delta(p_0+\epsilon)\Biggl]\Biggl\}.
\eea
Here the first part corresponds to the zero temperature case and the second and third parts are finite temperature effects.

\subsubsection{A gravitational scattering at finite temperature}

In the framework of the teleparallel gravity, there is a particular scattering process as given in FIG. 7.
\begin{figure}[h]
\includegraphics[scale=0.5]{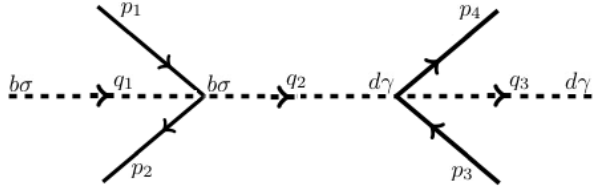}
\caption{A gravitational scattering}
\end{figure}
For this process the transition amplitude is
\begin{equation}
-i{\cal M}=\bar{u}(p_2){\cal V}^{b\sigma}u(p_1)D_{bd\sigma\gamma}(q_2)\bar{u}(p_4){\cal V}^{d\gamma}u(p_3),
\end{equation}
which at finite temperature yields
\bea
{\cal M}&=&\frac{e^2}{16}\bar{u}(p_2)\left(\gamma^\mu q_{1\mu}+q_{1\mu}\gamma^\mu\right)u(p_1)\Sigma_d\,^\sigma\Sigma^d_\sigma\,\bar{u}(p_4)\left(\gamma^\nu q_{3\nu}+q_{3\nu}\gamma^\nu\right)u(p_3)\times\nonumber\\
&\times &\left[\frac{\xi}{q_{20}^2-(\omega_{2q}-i\epsilon\xi)^2}+\frac{2\pi i\delta(q_{20}^2-\omega_{2q}^2)}{e^{\beta \omega_{2q}}-1}\left( \begin{array}{cc}1&e^{\beta \omega_{2q}/2}\\e^{\beta \omega_{2q}/2}&1\end{array} \right)\right].
\eea
The first term corresponds to the zero temperature results and the second part is finite temperature effects.

\section{Conclusions}

Transition amplitudes for interactions of fermions and gravitons are calculated at finite temperature. The gravitons are described by the teleparallel gravity in the weak field approximation, an alternative theory of gravitation which is formulated in a space with torsion. Teleparallel gravity provides a viable expression to deal with the interaction of gravitons and fermions since their coupling is due to the contorsion. The temperature of such processes is introduced by Thermo Field Dynamics which has the advantage of preserving the temporal coordinate. The scattering between gravitons and fermions at finite temperature may have been important in the formation of matter in the development of the universe when the temperature was higher and effects of quantum gravity were relevant.

\section*{Acknowledgments}

This work by A. F. S. is supported by CNPq projects 476166/2013-6 and 201273/2015-2. S. C. U. thanks the Funda\c{c}\~ao de Apoio $\grave{a}$ Pesquisa do Distrito Federal - FAPDF for financial support. We thank Physics Department, University of Victoria for access to facilities.


\end{document}